\documentclass[a4paper,11pt]{article}
\usepackage{latexsym}
\usepackage{amsmath}
\usepackage{amssymb}
\usepackage{amscd}
\newcommand{\sxi}[1]{{\rm sh}\left(\frac{1}{2\xi}(#1)\right)}

\newlength{\minitwocolumn}
\makeatletter
\@addtoreset{equation}{section}
\makeatother

\title{\bf 
Quantum Knizhnik-Zamolodchikov equation\\
associated with
$U_q(A_2^{(2)})$ for $|q|=1$
}

\author{Takeo Kojima}
\date{\it
Department of Mathematics,
College of Science and Technology,\\
Nihon University, Chiyoda-ku, Tokyo
101-0062, Japan\\
{\rm \today}
}

\begin{document}
\maketitle

\begin{abstract}
We present an integral representation to
the quantum Knizhnik-Zamolodchikov
equation associated
with twisted affine symmetry 
$U_q(A_2^{(2)})$ for massless regime $|q|=1$.
Upon specialization,
it leads to a cojectural formula
for the correlation function
of the Izergin-Korepin model in massless regime
$|q|=1$.
In a limiting case
$q \to -1$,
our conjectural formula
reproduce the correlation function
for the Izergin-Korepin model
\cite{HYZ, M}
at critical point $q=-1$.
\end{abstract}

\section{Introduction}
We shall
consider the quantum Knizhnik-Zamolodchikov
equation ($q$KZ equation) 
associated with twisted affine symmetry
$U_q(A_2^{(2)})$ for massless regime $|q|=1$.
In the earlier work
\cite{HYZ, M},
the $q$KZ equation associated with twisted affine symmetry
$U_q(A_2^{(2)})$ for massive regime $-1<{q}<0$,
was studied within
the framework of the representation theory
of $U_q(A_2^{(2)})$,
\cite{J, D}.
Hou et al. gave free field realizations
of the vertex operators,
and realized integral representations
of $q$KZ equation,
as the trace of the vertex operators.
Their integral representation
gave the correlation function
for the Izergin-Korepin model 
\cite{IK}
for massive regime 
$-1<{q}<0$.
In this paper we
present an integral representation
for $q$KZ equation associated with
twisted algebra $U_q(A_2^{(2)})$ for massless
regime $|q|=1$.
Let $V={\mathbb C}^3$,
and consider the $R$-matrix $R(\beta)
\in {\rm End}(V \otimes V)$
associated with $U_q(A_2^{(2)})$ (see sec.2).
The $q$KZ equation associated
with $U_q(A_2^{(2)})$
for $|q|=1$ is the following system of linear
difference equations for an unknown
function $G_{2N}(\beta_1,\cdots,\beta_{2N})$
taking value in the space $V^{\otimes 2N}$.
\begin{eqnarray}
&&G_{2N}
(\beta_1,\cdots,\beta_{j-1},\beta_j-\lambda i,
\beta_{j+1}, \cdots, \beta_{2N})\nonumber\\
&=&
R_{j,j+1}(\beta_j-\beta_{j+1}-\lambda i)^{-1}
\cdots R_{j,2N}
(\beta_j-\beta_{2N}-\lambda i)^{-1}
\nonumber\\
&\times&
R_{1,j}(\beta_1-\beta_j)\cdots
R_{j-1,j}(\beta_{j-1}-\beta_j)
G_{2N}
(\beta_1,\cdots,\beta_{j-1},\beta_j,
\beta_{j+1},\cdots,\beta_{2N}).
\end{eqnarray}
Here $R_{j,k}(\beta) \in {\rm End}
(V^{\otimes 2N})$
signifies the matrix acting as $R(\beta)$
on $(j,k)$-th tensor components and as identity elsewhere,
and diffence parameter $\lambda>0$.
Upon specialization of difference parameter
$\lambda=3\pi$
and spectral parameters
$\beta_j=\beta+3\pi i, \beta_{j+N
}=\beta, (1\leq j \leq N)$
, our integral representation
leads
to a conjectural formula
of the $N$-point correlation functions
of the Izergin-Korepin model
in massless regime $|q|=1$.
\begin{eqnarray}
G_{2N}(\beta+3\pi i, \cdots, \beta+3\pi i,
\beta,\cdots, \beta).
\end{eqnarray}
Actually, in a limiting case $q \to -1$,
our conjectural formula for
the correlation function reproduce
those of earlier work \cite{HYZ} at critical point
$q=-1$.

In connection with ``massless'' $q$KZ,
we should mention about the work \cite{S, JM1, JKM},
in which the authors presented
an integral representation to $q$KZ equation 
associated with $U_q(A_1^{(1)})$ for $|q|=1$.
In pioneering work \cite{S},
F.Smirnov gave conjectural integral representations
for the form factors of sine-Gordon model.
M.Jimbo et al.'s 
integral representation 
\cite{JM1} gave a conjectural formula
of the correlation function for the massless XXZ chain.
The higher spin (spin 1) generalization
of work \cite{JM1, JKM} was achieved in \cite{K1}. 
The $U_q(A_{n-1}^{(1)})$ generalization
of work \cite{JM1, JKM} was achieved in \cite{KY}.
T.Miwa and Y.Takeyama
\cite{MT} 
studied $q$KZ equation associated with
$U_q(A_{n-1}^{(1)})$ for $|q|=1$,
and
presented hypergeometric pairing 
in terminology of V.Tarasov and A.Varchenko
\cite{TV}.

Now a few words about organization
of this paper.
In section 2 we formulate 
the system of difference equations.
In section 3
we present an integral representation.
In section 4 we show that our integral representation
satisfies the system of diffence equations.
In section 5 we give a supporting argument
of our conjectural formula.

\section{Difference Equations}

In this section, we formulate the system of equations
we are going to study,
including the $q$KZ equation 
associated with $U_q(A_2^{(2)})$ for $|q|=1$.
In this paper we parametrize 
the deformation parameter $q$
of $U_q(A_2^{(2)})$ by $\xi$ as follows.
\begin{eqnarray}
q=-e^{-\frac{\pi i}{\xi}},~~\xi>1.
\end{eqnarray}
Consider the $R$-matrix $R(\beta) 
\in {\rm End}(V \otimes V)$
of $U_q(A_2^{(2)})$
acting on the tensor product of $V={\mathbb C}v_{-1}
\oplus {\mathbb C}v_0 \oplus {\mathbb C}v_1$ :
\begin{eqnarray}
R(\beta)v_{j_1}\otimes v_{j_2}&=&
\sum_{k_1,k_2=\pm 1,0}
v_{k_1}\otimes v_{k_2}
R(\beta)_{k_1 k_2}^{j_1 j_2},\\
R(\beta)&=&\frac{1}{\kappa(\beta)}\overline{R}(\beta).
\end{eqnarray}
The scalar function
$\kappa(\beta)$ will be specified below.
It is chosen to ensure that the $R$-matrix
satisfies the unitarity (\ref{unitary}) 
and the crossing symmetry (\ref{crossing}).
Nonzero entries of the $R$-matrix 
$\overline{R}(\beta)$ are given as follows.
\begin{eqnarray}
&&\overline{R}
(\beta)_{1,1}^{1,1}=
\overline{R}(\beta)_{-1,-1}^{-1,-1}
=1,
\nonumber\\
&&\overline{R}(\beta)_{0,-1}^{-1,0}
=\overline{R}(\beta)_{1,0}^{0,1}=
\frac{-e^{-\frac{\beta}{2\xi}}
{\rm sh}\left(\frac{\pi i}{\xi}\right)}{
{\rm sh}\left(\frac{1}{2\xi}(\beta-2\pi i)\right)
},\nonumber\\
&&
\overline{R}(\beta)_{-1,0}^{0,-1}=
\overline{R}(\beta)_{0,1}^{1,0}=
\frac{-e^{\frac{\beta}{2\xi}}
{\rm sh}\left(\frac{\pi i}{\xi}\right)}{
{\rm sh}\left(\frac{1}{2\xi}(\beta-2\pi i)\right)
},
\nonumber\\
&&\overline{R}(\beta)_{-1,0}^{-1,0}=
\overline{R}(\beta)_{0,-1}^{0,-1}=
\overline{R}(\beta)_{1,0}^{1,0}=
\overline{R}(\beta)_{0,1}^{0,1}=
\frac{-{\rm sh}\left(\frac{\beta}{2\xi}\right)}
{{\rm sh}\left(\frac{1}{2\xi}(\beta-2\pi i)\right)}
\nonumber\\
&&
\overline{
R}
(\beta)_{-1,1}^{-1,1}=
\overline{R}
(\beta)_{1,-1}^{1,-1}=
\frac{{\rm sh}
\left(\frac{\beta}{2\xi}\right)
{\rm sh}
\left(\frac{1}{2\xi}(\beta-\pi i)\right)
}{{\rm sh}
\left(\frac{1}{2\xi}(\beta-2\pi i)\right)
{\rm sh}
\left(\frac{1}{2\xi}(\beta-3\pi i)\right)
},\nonumber\\
&&
\overline{R}
(\beta)_{1,-1}^{-1,1}=
\frac{
{\rm sh}\left(\frac{\pi i}{\xi}\right)
\left(
-e^{-\frac{\pi i}{2\xi}}
{\rm ch}\left(\frac{\pi i}{\xi}\right)
+e^{\frac{-\beta+\pi i}{\xi}}
{\rm ch}\left(\frac{\pi i}{2\xi}\right)\right)
}{
{\rm sh}\left(\frac{1}{2\xi}(\beta-2\pi i)\right)
{\rm sh}\left(\frac{1}{2\xi}(\beta-3\pi i)\right)
},\nonumber\\
&&\overline{R}
(\beta)_{-1,1}^{1,-1}=
\frac{
{\rm sh}\left(\frac{\pi i}{\xi}\right)
\left(
e^{\frac{\pi i}{2\xi}}
{\rm ch}\left(\frac{\pi i}{\xi}\right)
-e^{\frac{\beta-\pi i}{\xi}}
{\rm ch}\left(\frac{\pi i}{2\xi}\right)\right)
}{
{\rm sh}\left(\frac{1}{2\xi}(\beta-2\pi i)\right)
{\rm sh}\left(\frac{1}{2\xi}(\beta-3\pi i)\right)
},\nonumber\\
&&
\overline{R}(\beta)_{0,0}^{1,-1}=
\overline{R}(\beta)_{-1,1}^{0,0}=\frac{
i e^{\frac{\beta-2\pi i}{2\xi}}
{\rm sh}\left(\frac{\pi i}{\xi}\right)
{\rm sh}\left(\frac{\beta}{2\xi}\right)
}{
{\rm sh}\left(\frac{1}{2\xi}(\beta-2\pi i)\right)
{\rm sh}\left(\frac{1}{2\xi}(\beta-3\pi i)\right)
},
\nonumber
\\
&&\overline{R}(\beta)_{0,0}^{-1,1}=
\overline{R}(\beta)_{1,-1}^{0,0}=
\frac{-i e^{\frac{-\beta+2\pi i}{2\xi}}
{\rm sh}\left(\frac{\pi i}{\xi}\right)
{\rm sh}\left(\frac{\beta}{2\xi}\right)
}{
{\rm sh}\left(\frac{1}{2\xi}(\beta-2\pi i)\right)
{\rm sh}\left(\frac{1}{2\xi}(\beta-3\pi i)\right)
},\nonumber\\
&&\overline{R}(\beta)_{0,0}^{0,0}=-
\frac{{\rm sh}\left(\frac{\beta}{2\xi}\right)}{{\rm sh}
\left(\frac{1}{2\xi}(\beta-2\pi i)\right)}
-
\frac{
{\rm sh}\left(\frac{\pi i}{\xi}\right)
{\rm sh}\left(\frac{3\pi i}{2\xi}\right)
}{
{\rm sh}\left(\frac{1}{2\xi}(\beta-2\pi i)\right)
{\rm sh}\left(\frac{1}{2\xi}(\beta-3\pi i)\right)
}.
\end{eqnarray}
Our $R$-matrix $R(\beta)$
gives the Boltzmann weight of the Izergin-Korepin
model \cite{IK}.
The $R$-matrix $R(\beta)$ satisfies
the Yang-Baxter equation.
\begin{eqnarray}
R_{12}(\beta_1-\beta_2)
R_{13}(\beta_1-\beta_3)R_{23}(\beta_2-\beta_3)=
R_{23}(\beta_2-\beta_3)
R_{13}(\beta_1-\beta_3)R_{12}(\beta_1-\beta_2),
\end{eqnarray}
the unitarity condition, 
\begin{eqnarray}
R_{12}(\beta)R_{21}(-\beta)&=&id,\label{unitary}
\end{eqnarray}
and the crossing symmetry,
\begin{eqnarray}
R\left(\beta+\frac{3\pi i}{2}\right)_{j_1,j_2}^{k_1,k_2}
&=&(ie^{-\frac{\pi i}{2\xi}})^{k_2-j_2}\times
R\left(-\beta+\frac{3\pi i}{2}\right)_{-k_2,j_1}^{-j_2,k_1}.
\label{crossing}
\end{eqnarray}
Let $N$ be a non-negative integer.
Consider a $V^{\otimes 2N}$-valued function
$G_{2N}(\beta_1,\cdots,\beta_{2N})$,
depending on the 'spectral parameter'
$\beta_1, \cdots, \beta_{2N}$.
\begin{eqnarray}
G_{2N}(\beta_1,\cdots,\beta_{2N})=
\sum_{j_1,\cdots,j_{2N}=\pm1,0}
v_{j_1}\otimes \cdots \otimes v_{j_{2N}}
G_{2N}
(\beta_1,\cdots,\beta_{2N})
_{j_1,\cdots,j_{2N}}.
\end{eqnarray}
We study the following system of difference equations
for $G_{2N}
(\beta_1,\cdots,\beta_{2N})
_{j_1,\cdots,j_{2N}}$
involving parameter $\lambda$.
Hereafter we assume that
\begin{eqnarray}
2(\xi+1)\pi>\lambda>0,~~\xi>1.
\end{eqnarray}
{\bf $R$-matrix Symmetry}
\begin{eqnarray}
&&G_{2N}
(\beta_1,\cdots,\beta_{s+1},\beta_s,\cdots,\beta_{2N})
_{j_1,\cdots,j_{s+1},j_s,
\cdots,j_{2N}}\nonumber\\
&=&
\sum_{j_s',j_{s+1}'=\pm1,0}
R(\beta_s-\beta_{s+1})_{j_s,j_{s+1}}^{j_s'
,j_{s+1}'}
G_{2N}
(\beta_1,\cdots,\beta_{s},\beta_{s+1},\cdots,\beta_{2N})
_{j_1,\cdots,j_s',j_{s+1}',\cdots,j_{2N}}.
\label{Dif1}
\end{eqnarray}
{\bf Cyclicity Condition}
\begin{eqnarray}
G_{2N}
(\beta_1,\cdots,\beta_{2N-1},\beta_{2N}
-i\lambda)
_{j_1,\cdots,j_{2N}}
=G_{2N}(\beta_{2N}
,\beta_2,\cdots,\beta_{2N-1})
_{j_{2N},j_1,\cdots,j_{2N-1}}.
\label{Dif2}
\end{eqnarray}
{\bf Recursion Relation}
\begin{eqnarray}
&&G_{2N}
(\beta_1,\cdots,\beta_{s-1},\beta,\beta+3\pi i,
\beta_{s+1},\cdots,\beta_{2N})
_{j_1,\cdots,j_{s-1},j,-j,
j_{s+2},\cdots,j_{2N}}\nonumber\\
&=&C_{j}
~G_{2N-2}(\beta_1,\cdots,\beta_{s-1},
\beta_{s+1},\cdots,\beta_{2N})
_{j_1,\cdots,j_{s-1},
j_{s+2},\cdots,j_{2N}
}.\label{Dif3}
\end{eqnarray}
Here $C_j$ are specified below. 
See (\ref{C1}), (\ref{C0}) and (\ref{C-1}).
Note that the above system of equations
involve only the functions
$G_{2N}
(\beta_1,\cdots,\beta_{2N})
_{j_1,\cdots,j_{2N}}$
with fixed value of 'spin'
$j_1+\cdots+j_{2N}=0$.

\section{Integral Representation}

The purpose of this section is
to 
present an integral representation.
In what follows we use the Multiple-Gamma functions
$\Gamma_r(x|\omega_1 \cdots \omega_r)$
and the Multiple-Trigonometric functions
$S_r(x|\omega_1 \cdots \omega_r)$
introduced in \cite{JM1} as follows.
\begin{eqnarray}
{\rm log}\Gamma_r(x|\omega_1 \cdots \omega_r)=
\gamma \frac{(-1)^r}{r!}B_{r,r}(x|\omega_1,\cdots,
\omega_r)+\int_{C'}
\frac{e^{-xt}{\rm log}(-t)}{
\prod_{j=1}^r (1-e^{-\omega_j t})}
\frac{dt}{2\pi i t},~({\rm Re}x>0),
\end{eqnarray}
where $\gamma$=Euler's constant,
the integral contour $C'$ is shown in below Figure
(Contour $C'$),
and
Multiple-Bernoulli polynomials
$B_{r,r}(x|\omega_1 \cdots \omega_r)$
are given by 
\begin{eqnarray}
\frac{t^r e^{xt}}{\prod_{j=1}^r
(e^{\omega_j t}-1)}=
\sum_{n=0}^\infty
\frac{t^n}{n!}B_{r,n}(x|\omega_1,\cdots,\omega_r).
\end{eqnarray}
The Multiple-Gamma function
$\Gamma_r(x|\omega_1 \cdots \omega_r)$
is an entire function of $x$.
$\Gamma_r(x|\omega_1 \cdots \omega_r)$
is meromorphic function
with poles at
$x=n_1\omega_1+\cdots+
n_r\omega_r,
(n_1,\cdots,n_r \leq 0)$.\\
Let us set Multiple-Trigonometric function by
\begin{eqnarray}
S_r(x|\omega_1,\cdots,\omega_r)=
\Gamma_r(x|\omega_1,\cdots,\omega_r)^{-1}
\Gamma_r(\omega_1+\cdots+\omega_r-x|
\omega_1,\cdots,\omega_r)^{(-1)^r}.
\end{eqnarray}
Properties of these functions are listed in \cite{JM1}.
For examples they enjoy
\begin{eqnarray}
\frac{\Gamma_{r}(x+\omega_r|\omega_1,\cdots,\omega_{r})}{
\Gamma_{r}(x|\omega_1,\cdots,\omega_{r})
}=
\frac{1}{\Gamma_{r-1}(x|\omega_1,\cdots,\omega_{r-1})},\\
\frac{
S_{r}(x+\omega_r|\omega_1,\cdots,
\omega_{r})}{S_{r}(x|\omega_1,\cdots,
\omega_{r})}=\frac{1}{S_{r-1}(x|\omega_1,\cdots,
\omega_{r-1})},
\end{eqnarray}
and
\begin{eqnarray}
\Gamma_1(x|\omega)=
\omega^{\frac{x}{\omega}-\frac{1}{2}}
\frac{\Gamma(x/\omega)}{\sqrt{2\pi}},~
S_1(x|\omega)=2{\rm sin}\left(\frac{\pi x}{\omega}\right).
\end{eqnarray}
Here $\Gamma(x)$ is Gamma function.
As $x \to \infty ~(\pm{\rm Im}x>0)$, the function
$S_2(x|\omega_1 \omega_2)$ behaves as follows.
\begin{eqnarray}
{\rm log}S_2(x|\omega_1 \omega_2)=
\pm \pi i
\left(
\frac{x^2}{2\omega_1 \omega_2}-\frac{\omega_1+
\omega_2}{2\omega_1 \omega_2}x-\frac{1}{12}
\left(\frac{\omega_1}{\omega_2}+
\frac{\omega_2}{\omega_1}+3\right)\right)+o(1).
\label{asymptotic}
\end{eqnarray}
\\
~\\
~\\
%WinTpicVersion2.13
\unitlength 0.1in
\begin{picture}(34.10,11.35)(17.90,-19.35)
% VECTOR 2 0 3 0
% 4 5200 1200 2190 1200 2190 2000 5190 2000
% 
\special{pn 8}%
\special{pa 5200 800}%
\special{pa 2190 800}%
\special{fp}%
\special{sh 1}%
\special{pa 2190 800}%
\special{pa 2257 820}%
\special{pa 2243 800}%
\special{pa 2257 780}%
\special{pa 2190 800}%
\special{fp}%
\special{pa 2190 1600}%
\special{pa 5190 1600}%
\special{fp}%
\special{sh 1}%
\special{pa 5190 1600}%
\special{pa 5123 1580}%
\special{pa 5137 1600}%
\special{pa 5123 1620}%
\special{pa 5190 1600}%
\special{fp}%
% LINE 2 0 3 0
% 2 5190 1600 2590 1610
% 
\special{pn 8}%
\special{pa 5190 1200}%
\special{pa 2590 1210}%
\special{fp}%
% STR 2 0 3 0
% 3 2590 1510 2590 1610 2 0
% $0$
\put(25.9000,-12.1000){\makebox(0,0)[lb]{$0$}}%
% CIRCLE 2 0 3 0
% 4 2190 1610 2190 1210 2190 1210 2190 4410
% 
\special{pn 8}%
\special{ar 2190 1210 400 400  1.5707963 4.7123890}%
% STR 2 0 3 0
% 3 3390 2320 3390 2420 5 0
% {\bf Contour} $C'$
\put(33.9000,-20.2000){\makebox(0,0){{\bf Contour} $C'$}}%
\end{picture}%

\subsection{Auxiliary functions}
The integral formula involves certain special
functions $\kappa(\beta),
\rho(\beta),
\varphi(\beta),
\psi(\beta)$.
\\
${\bf \bullet}~\kappa(\beta)$ :
Let us set
\begin{eqnarray}
\kappa(\beta)=
\frac{S_2(-i\beta+2\pi)
S_2(-i\beta+3\pi)
S_2(i\beta+5\pi)
S_2(i\beta+6\pi)
}{
S_2(i\beta+2\pi)
S_2(i\beta+3\pi)
S_2(-i\beta+5\pi)
S_2(-i\beta+6\pi)
},~(S_2(x)=S_2(x|2\pi\xi,6\pi)).~
\end{eqnarray}
The function $\kappa(\alpha)$
satisfies the following difference equations,
which ensure the unitarity and the crossing
symmetry of $R$-matrix.
\begin{eqnarray}
\kappa(\beta)\kappa(-\beta)&=&1,\\
\kappa(\beta)\kappa(\beta-3\pi i)&=&
\frac{\sxi{\beta-\pi i}
{\rm sh}\left(\frac{\beta}{2\xi}\right)}
{\sxi{\beta-2\pi i}\sxi{\beta-3\pi i}}.
\end{eqnarray}
${\bf \bullet}~\rho(\beta)$~:~
Let us set
\begin{eqnarray}
\rho(\beta)=
\frac
{
S_3(-i\beta+2\pi)
S_3(-i\beta+3\pi)
S_3(i\beta+3\pi+\lambda)
S_3(i\beta+2\pi+\lambda)
}
{S_3(-i\beta+2\pi\xi)
S_3(-i\beta+\pi+2\pi\xi)
S_3(i\beta+\pi+2\pi\xi+\lambda)
S_3(i\beta+2\pi\xi+\lambda)},
\label{rho}
\\
(S_3(x)=S_3(x|2\pi\xi,\lambda,6\pi)).
~~~~~~~~~~~~~~~~~~~~~~~~~~~~\nonumber
\end{eqnarray}
The function $\rho(\alpha)$ satisfies
the following difference equations.
\begin{eqnarray}
\frac{\rho(\beta)}{\rho(-\beta)}=\kappa(\beta),
~~\rho(i\lambda-\beta)=\rho(\beta).
\end{eqnarray}
${\bf \bullet}~\varphi(\alpha)$~:~
Let us set
\begin{eqnarray}
\varphi(\alpha)&=&\frac{1}{
S_2(i\alpha+\pi|\lambda,2\pi\xi)
S_2(-i\alpha+\pi|\lambda,2\pi\xi)}.
\label{kernel1}
\end{eqnarray}
The function $\varphi(\alpha)$
satisfies following difference equations.
\begin{eqnarray}
\varphi(\alpha)&=&\varphi(-\alpha),\\
\frac{\varphi(\alpha-i\lambda)}{
\varphi(\alpha)}&=&-
\frac
{
\sxi{\alpha-\pi i}
}
{\sxi{\alpha+\pi i-\lambda i}
},
\\
\frac{\varphi(\alpha \pm 2\pi\xi i)}{
\varphi(\alpha)}&=&-
\frac
{
\displaystyle
{\rm sh}\left(
\frac{\pi}{\lambda}(\alpha\pm \pi i)\right)}
{\displaystyle
{\rm sh}\left(\frac{\pi}{\lambda}(
\alpha\mp\pi i \pm 2\pi\xi i)\right)}.
\end{eqnarray}
The function $\varphi(\alpha)$
has poles at
\begin{eqnarray}
\varphi(\alpha)=\pm i
\left(
n_1\lambda+n_2 2\pi\xi+\pi\right),~~n_1,n_2\geq 0.
\end{eqnarray}
The function $\varphi(\alpha)$
is evaluated as follows.
\begin{eqnarray}
|\varphi(\alpha)|&\leq& Const.
\exp\left(
\frac{2\pi-(\lambda+2\pi\xi)}{2\xi\lambda}
|\alpha|\right),~|\alpha|\to \infty,\\
\varphi(\alpha)&=&
\frac{\sqrt{\frac{\lambda \xi}{2\pi}}}{i
S_2(2\pi|\lambda,2\pi\xi)}
\times
\frac{1}{\alpha-\pi i}+\cdots,~\alpha
\to \pi i.
\end{eqnarray}
${\bf \bullet}~\psi(\alpha)$~:~
Let us set
\begin{eqnarray}
\psi(\alpha)&=&\frac{1}{S_2(i\alpha-2\pi|\lambda,2\pi\xi)
S_2(-i\alpha-2\pi|\lambda,2\pi\xi)},
\label{kernel2}
\end{eqnarray}
The function $\psi(\alpha)$ satisfies
the following difference equations.
\begin{eqnarray}
\psi(\alpha)&=&\psi(-\alpha),\\
\frac{\psi(\alpha-i\lambda)}{\psi(\alpha)}&=&
-\frac
{\sxi{\alpha+2\pi i}}
{\sxi{\alpha-\lambda i-2\pi i}},
\\
\frac{\psi(\alpha\pm 2\pi\xi i)}{
\psi(\beta)}&=&-
\frac
{
\displaystyle
{\rm sh}\left(\frac{\pi}{\lambda}(\alpha \mp 2\pi i)\right)
}
{\displaystyle
{\rm sh}\left(\frac{\pi}{\lambda}(\alpha \pm 2\pi i
\pm 2\pi \xi i)\right)
},
\end{eqnarray}
and
\begin{eqnarray}
&&\varphi(\alpha+\pi i)
\psi(\alpha)
\varphi(\alpha-\pi i)\\
&=&2^{-6}
\left\{
\sxi{\alpha+2\pi i}
{\rm sh}\left(\frac{\alpha}{2\xi}\right)
\sxi{-\alpha+2\pi i}\right\}^{-1}\nonumber\\
&\times&
\left\{
{\rm sh}\left(\frac{\pi}{\lambda}(
\alpha-2\pi i)\right)
{\rm sh}\left(\frac{\pi}{\lambda}\alpha
\right)
{\rm sh}\left(\frac{\pi}{\lambda}(
\alpha+\pi i)\right)
\right\}^{-1}.
\end{eqnarray}
The function $\psi(\alpha)$ has poles at
\begin{eqnarray}
\alpha=\pm i \left(n_1\lambda+
n_2 2\pi\xi-2\pi\right),~~n_1,n_2 \geq 0.
\end{eqnarray}
The function $\psi(\alpha)$
is evaluated as follows.
\begin{eqnarray}
|\psi(\alpha)|\leq Const.
\exp\left(-\frac{4\pi+\lambda+2\pi\xi}
{2\xi\lambda}|\alpha|\right),~|\alpha|\to
\infty.
\end{eqnarray}

\subsection{Integral Representation}
In this section
we present the integral representation
for
$G_{2N}
(\beta_1,\cdots,\beta_{2N})
_{j_1,\cdots,j_{2N}}$,
which is our main result of this paper.
At first we demonstrate
integral representations for
$N=1$ case,
$G_2(\beta_1,\beta_2)_{0,0},
G_2(\beta_1,\beta_2)_{1,-1}$,
and $G_2(\beta_1,\beta_2)_{-1,1}$.

\begin{eqnarray}
&&G_2(\beta_1,\beta_2)_{0,0}=
2^9
e^{-\frac{\pi i}{\xi}}{\rm sh}\left(\frac{\pi i}{\xi}\right)
\rho(\beta_1-\beta_2)
\int_{C}
\frac{d\alpha_1}{2\pi i}
\int_{C}
\frac{d\alpha_2}{2\pi i}\nonumber\\
&\times&
\left\{\varphi(\alpha_1-\alpha_2)
\psi(\alpha_1-\alpha_2)
\prod_{s,t=1,2}
\varphi(\alpha_s-\beta_t)\right\}
e^{\frac{1}{2\xi}(\alpha_1+\alpha_2
-\beta_1-\beta_2)}
h_\lambda(\alpha_1-\alpha_2)\\
&\times&
\sxi{\alpha_1-\alpha_2+2\pi i}
\sxi{-\alpha_1+\alpha_2+\pi i}
\sxi{\alpha_2-\beta_1+\pi i}
\sxi{-\alpha_1+\beta_2+\pi i}.\nonumber
\end{eqnarray}

\begin{eqnarray}
&&G_2(\beta_1,\beta_2)_{-1,1}=
2^9
ie^{-\frac{\pi i}{\xi}}{\rm sh}\left(\frac{\pi i}{\xi}\right)
\rho(\beta_1-\beta_2)
\int_{C}
\frac{d\alpha_1}{2\pi i}
\int_{C}
\frac{d\alpha_2}{2\pi i}\nonumber\\
&\times&
\left\{\varphi(\alpha_1-\alpha_2)
\psi(\alpha_1-\alpha_2)
\prod_{s,t=1,2}
\varphi(\alpha_s-\beta_t)\right\}
e^{-\frac{1}{2\xi}\beta_2}
h_\lambda(\alpha_1-\alpha_2)\\
&\times&
\sxi{\alpha_1-\alpha_2+2\pi i}
\sxi{-\alpha_1+\alpha_2+\pi i}
\sxi{\alpha_1-\beta_1+\pi i}
\sxi{\alpha_2-\beta_1+\pi i}\nonumber\\
&\times&
\left(
e^{\frac{1}{2\xi}\alpha_1}
\sxi{\alpha_2-\beta_2+\pi i}
-e^{\frac{1}{2\xi}\alpha_2}
\sxi{\beta_2-\alpha_1+\pi i}
\right),\nonumber
\end{eqnarray}
and
\begin{eqnarray}
&&G_2(\beta_1,\beta_2)_{1,-1}=2^9
ie^{-\frac{\pi i}{\xi}}{\rm sh}\left(\frac{\pi i}{\xi}\right)
\rho(\beta_1-\beta_2)
\int_{C}
\frac{d\alpha_1}{2\pi i}
\int_{C}
\frac{d\alpha_2}{2\pi i}\nonumber\\
&\times&
\left\{\varphi(\alpha_1-\alpha_2)
\psi(\alpha_1-\alpha_2)
\prod_{s,t=1,2}
\varphi(\alpha_s-\beta_t)\right\}
e^{-\frac{1}{2\xi}\beta_1}
h_\lambda(\alpha_1-\alpha_2)\\
&\times&
\sxi{\alpha_1-\alpha_2+2\pi i}
\sxi{-\alpha_1+\alpha_2+\pi i}
\sxi{-\alpha_1+\beta_2+\pi i}
\sxi{-\alpha_2+\beta_2+\pi i}\nonumber\\
&\times&
\left(
e^{\frac{1}{2\xi}\alpha_1}
\sxi{\alpha_2-\beta_1+\pi i}
-e^{\frac{1}{2\xi}\alpha_2}
\sxi{\beta_1-\alpha_1+\pi i}
\right),\nonumber
\end{eqnarray}
Here we have set
\begin{eqnarray}
h_\lambda(\alpha)=
{\rm sh}\left(\frac{\pi}{\lambda}(\alpha-2\pi i)\right)
{\rm sh}\left(\frac{\pi}{\lambda}\alpha\right)
{\rm sh}\left(\frac{\pi}{\lambda}(\alpha+2\pi i)\right).\label{h}
\end{eqnarray}
The integral contour $C$ is taken 
along a path going from $-\infty$
to $\infty$
in such a way that
$-\pi<{\rm Im}(\alpha_k-\beta_j)<\pi$
for all $k,j$. 

Next we present 
the integral representation for 
$G_{2N}
(\beta_1,\cdots,\beta_{2N})
_{j_1,\cdots,j_{2N}}$.
This is {\bf Main Result} of this paper.
Given a set of indices $j_1,\cdots,j_{2N} \in
\{-1,0,1\}$.
Let us set index set ${\cal A}_j,~(j=-1,0,1)$ by
\begin{eqnarray}
{\cal A}_{j}=\{1\leq s \leq 2N|
j_s \geq j\}.
\end{eqnarray}
The index set ${\cal A}_{-1}=\{1,\cdots,2N\}$.
Let us set $\alpha_{-1,s}=\beta_s, (1\leq s 
\leq 2N)$.
Let us define the kernel function 
$\Psi(\{\alpha\}|\{\beta\})$
by
\begin{eqnarray}
\Psi(\{\alpha\}|\{\beta\})&=&
\prod_{s=1}^{2N}
\prod_{j=0,1}\prod_{t \in {\cal A}_j}
\varphi(\beta_s-\alpha_{j,t})
\prod_{s \in {\cal A}_0,
t \in {\cal A}_1}
\varphi(\alpha_{0,s}-\alpha_{1,t})
\psi(\alpha_{0,s}-\alpha_{1,t})\nonumber\\
&\times&
\prod_{j=0,1}\prod_{s,t \in {\cal A}_j
\atop{s<t}}
\varphi(\alpha_{j,s}-\alpha_{j,t})
\psi(\alpha_{j,s}-\alpha_{j,t}).
\label{kernel3}
\end{eqnarray}
Here we have use
the function
$\varphi(\alpha), \psi(\alpha)$
in (\ref{kernel1}), (\ref{kernel2}).
Let us define the auxiliary functions
$g_k(\{\alpha_{j}\}_{j=-1}^{k}), (k=\pm1,0)$ 
by
\begin{eqnarray}
&&g_{-1}(\{\alpha_{-1}\})=1,
\label{g1}\\
&&g_0(\{\alpha_{-1},\alpha_0\})=
\frac{e^{\frac{1}{2\xi}(\alpha_0-\alpha_{-1})}
{\rm sh}\left(\frac{\pi i}{\xi}\right)}
{\sxi{\alpha_0-\alpha_{-1}+\pi i}},
\label{g2}\\
&&g_1(\{\alpha_{-1},\alpha_0,\alpha_1\})=
\frac{ie^{-\frac{\pi i}{\xi}}
{\rm sh}\left(\frac{\pi i}{\xi}\right)}{
\sxi{\alpha_0-\alpha_{-1}+\pi i}
\sxi{\alpha_1-\alpha_{-1}+\pi i}
}\nonumber\\
&\times&
\left(
e^{\frac{1}{2\xi}\alpha_0}
\sxi{\alpha_1-\alpha_{-1}+\pi i}
-
e^{\frac{1}{2\xi}\alpha_1}
\sxi{\alpha_{-1}-\alpha_{0}+\pi i}
\right)
.\label{g3}
\end{eqnarray}
Let us set the integral representation
$G_{2N}$ by
\begin{eqnarray}
&&G_{2N}
(\beta_1,\cdots,\beta_{2N})
_{j_1,\cdots,j_{2N}}\nonumber
\\
&=&
2^{N(14N-5)}
\prod_{1\leq s<t \leq 2N}
\rho(\beta_s-\beta_t)
\prod_{s=1}^{2N}
\prod_{j=0}^{j_s}
\int_{C}
\frac{
d\alpha_{j,s}}{2\pi i}
\Psi(\{\alpha\}|\{\beta\})\nonumber\\
&\times&
\prod_{j=0,1}
\prod_{s,t \in {\cal A}_j
\atop{s<t}}
h_\lambda(\alpha_{j,s}-\alpha_{j,t})
\prod_{s \in {\cal A}_0, t \in {\cal A}_{1}
\atop{s \leq t}}
h_\lambda(\alpha_{0,s}-\alpha_{1,t})
\prod_{s \in {\cal A}_0, t \in {\cal A}_{1}
\atop{s > t}}
h_\lambda(\alpha_{1,t}-\alpha_{0,s})\nonumber\\
&\times&
\prod_{s=1}^{2N}
g_{j_s}(\{\alpha_{j,s}\}_{j=-1}^{j_s}) 
\prod_{j=0,1}
\prod_{s,t \in {\cal A}_j
\atop{s<t}}
\sxi{\alpha_{j,s}-\alpha_{j,t}+2\pi i}
\sxi{-\alpha_{j,s}+\alpha_{j,t}+\pi i
}\nonumber\\
&\times&
\prod_{s \in {\cal A}_0, t \in {\cal A}_{1}
\atop{s \leq t}}
\sxi{\alpha_{0,s}-\alpha_{1,t}+2\pi i}
\sxi{-\alpha_{0,s}+\alpha_{1,t}+\pi i
}\nonumber\\
&\times&
\prod_{s \in {\cal A}_0, t \in {\cal A}_{1}
\atop{s > t}}
\sxi{\alpha_{1,t}-\alpha_{0,s}+2\pi i}
\sxi{-\alpha_{1,t}+\alpha_{0,s}+\pi i}
\nonumber\\
&\times&
\prod_{s=1}^{2N}
\prod_{j=0,1}
\prod_{t \in {\cal A}_j
\atop{s \leq t}}
\sxi{-\beta_s+\alpha_{j,t}+\pi i}
\prod_{t \in {\cal A}_j
\atop{t<s}}
\sxi{-\alpha_{j,t}+\beta_s+\pi i}.\label{Main}
\end{eqnarray}
Here we have used 
$\rho(\beta)$ in (\ref{rho}),
$\Psi(\{\alpha\}|\{\beta\})$
in (\ref{kernel3}),
$h_\lambda(\alpha)$ in (\ref{h}),
$g_{j_s}(\{\alpha_{j,s}\}_{j=-1}^{j_s})$ in
(\ref{g1}), (\ref{g2}), (\ref{g3}).
Here the integral contour $C$
is taken along a path going from $-\infty$
to $\infty$ in such a way that
$-\pi<{\rm Im}(\alpha_k-\beta_j)<\pi$
for all $k,j$.
We understand $G_0=1$.
~\\

Let us check the convergence of the integral
of $G_{2N}$.
Let us set
\begin{eqnarray}
I_\lambda(\alpha)&=&
\varphi(\alpha)\psi(\alpha)
h_\lambda(\alpha)
\sxi{\alpha+2\pi i}\sxi{-\alpha+\pi i},
\label{f1}\\
J_\lambda(\alpha)&=&
\varphi(\alpha)
\sxi{-\alpha+\pi i}.
\label{f2}
\end{eqnarray}
Let us consider
the evaluation of $|\alpha_{k,t}|\to
\infty$.
In our integral representation,
factors involving
the variable $\alpha_{k,t}$
consist of following three parts.
\begin{eqnarray}
&&g_{j_t}(\{\alpha_{j,t}
\}_{j=-1}^{j_t})
\nonumber\\
&\times&
\prod_{s \leq t}
J_\lambda(-\beta_s+\alpha_{k,t})
\prod_{t<s}
J_\lambda(-\alpha_{k,t}+\beta_s)
\prod_{j=0,1}\\
&\times&
\prod_{s<t}
I_\lambda(\alpha_{j,s}-\alpha_{k,t})
\prod_{j=0,1}
\prod_{t<s}
I_\lambda(\alpha_{k,t}-\alpha_{j,s})
\prod_{j \neq k}
I_\lambda({\rm sgn}(k-j)
(\alpha_{j,t}-\alpha_{k,t})).
\nonumber
\end{eqnarray}
For $|\alpha| \to \infty$,
the factors $I_\lambda(\alpha),
J_\lambda(\alpha)$ are evaluated as follows.
\begin{eqnarray}
&&I_\lambda(\alpha)
=Const.\exp\left(
\frac{\pi}{\xi \lambda}(\xi-1)|\alpha|
\right),
\label{e1}\\
&&
J_\lambda(\alpha)
=Const.\exp\left(
\frac{\pi}{\xi \lambda}(1-\xi)|\alpha|
\right).
\label{e2}
\end{eqnarray}
We have
$g_{j_t}(\{\alpha_{j,t}\}_{j=-1}^{j_t})
\to 1,~(\alpha_{k,t} \to \infty)$.
Let us consider the evaluation for variable $
|\alpha_{k,t}|\to \infty$ in the
integrand of our integral formula.
Because of spin condition,
$j_1+\cdots+j_{2N}=0$,
there exist
$2N$ factors of type (\ref{f2}),
$(2N-1)$ factors of type (\ref{f1}),
and one $g_{j_t}(\{\alpha_{j,t}\}
_{j=1}^{j_t})$, in the integrand.
Using above estimates 
(\ref{e1}) (\ref{e2}),
we know convergence of
the integral representation upon 
condition $\xi>1$.

\section{Proof of Difference Equations}

In this section we prove
the integral formula
(\ref{Main}) satisfies
the system of difference equations
(\ref{Dif1}), (\ref{Dif2}), and (\ref{Dif3}).

\subsection{$R$-matrix Symmetry}

In this section we prove $R$-matrix symmetry
(\ref{Dif1}).
Let us set
\begin{eqnarray}
\overline{G}_{2N}
(\beta_1,\cdots,\beta_{2N})
_{j_1,\cdots,j_{2N}}=
\prod_{1\leq s<t\leq 2N}
\rho(\beta_s-\beta_t)^{-1}
{G}_{2N}
(\beta_1,\cdots,\beta_{2N})
_{j_1,\cdots,j_{2N}}.
\end{eqnarray}
Because of the properties of 
$\frac{\rho(\beta)}{\rho(-\beta)}=\kappa(\beta),
\rho(i\lambda-\beta)=\rho(\beta)$,
the equation (\ref{Dif1}) is reduced to the same equation
for $\overline{G}_{2N}$ 
where $R$ is replaced to $\overline{R}$.
\begin{eqnarray}
&&\overline{G}_{2N}(\beta_1,\cdots,
\beta_{s+1},\beta_s,\cdots,\beta_{2N})
_{j_1,\cdots,j_{s+1},j_s,\cdots,j_{2N}}
\nonumber\\
&=&
\sum_{j_s',j_{s+1}'=\pm 1,0}
\overline{R}(\beta_s-\beta_{s+1})_{j_s,j_{s+1}}
^{j_s',j_{s+1}'}
\overline{G}_{2N}(\beta_1,\cdots,
\beta_{s},\beta_{s+1},\cdots,\beta_{2N})
_{j_1,\cdots,j_s',j_{s+1}',\cdots,j_{2N}}.
\label{Dif1'}
\end{eqnarray}
${\bf \bullet}~N=1$ Case.\\
First we demonstrate how to prove (\ref{Dif1}) 
in simple cases, $N=1$.
There exist $3$ cases to consider.
We prove (\ref{Dif1}) by checking every cases.
Let us consider the case,
\begin{eqnarray}
&&\overline{G}_2(\beta_2,\beta_1)_{-1,1}
\label{G(-1,1)}.
\\
&=&
\overline{R}(\beta_1-\beta_2)_{1,-1}^{1,-1}
\overline{G}_2(\beta_1,\beta_2)_{1,-1}
+
\overline{R}(\beta_1-\beta_2)_{1,-1}^{0,0}
\overline{G}_2(\beta_1,\beta_2)_{0,0}
+
\overline{R}(\beta_1-\beta_2)_{1,-1}^{-1,1}
\overline{G}_2(\beta_1,\beta_2)_{-1,1}.\nonumber
\end{eqnarray}
Considering the integrand of (LHS)-(RHS),
we get
\begin{eqnarray}
2^9
e^{-\frac{\pi i}{\xi}}{\rm sh}\left(\frac{\pi i}{\xi}\right)
\int_{C}
\int_{C}
\frac{d\alpha_1}{2\pi i}
\frac{d\alpha_2}{2\pi i}
\left\{\psi(\alpha_1-\alpha_2)
\varphi(\alpha_1-\alpha_2)\prod_{s,t=1,2}
\varphi(\alpha_s-\beta_t)
\right\}\nonumber\\
\times h_\lambda(\alpha_1-\alpha_2)~
{\rm Int}(\alpha_1 \alpha_2|\beta_1 \beta_2)_{-1,1},
\end{eqnarray}
where we set
\begin{eqnarray}
&&{\rm Int}(\alpha_1 \alpha_2|\beta_1 \beta_2)_{-1,1}
\\
&=&
\sxi{\alpha_1-\alpha_2+2\pi i}
\sxi{-\alpha_1+\alpha_2+\pi i}\nonumber\\
&\times&
\left\{
i\sxi{-\alpha_1+\beta_1+\pi i}
\sxi{-\alpha_2+\beta_1+\pi i}
e^{-\frac{1}{2\xi}\beta_2}
\right.\nonumber\\
&\times&
\left(e^{\frac{1}{2\xi}\alpha_1}
\sxi{\alpha_2-\beta_2+\pi i}
-
e^{\frac{1}{2\xi}\alpha_2}
\sxi{\beta_2-\alpha_1+\pi i}
\right)
\nonumber\\
&-&
\overline{R}(\beta_1-\beta_2)_{1,-1}^{1,-1}
i \sxi{-\alpha_1+\beta_2+\pi i}
\sxi{-\alpha_2+\beta_2+\pi i}
e^{-\frac{1}{2\xi}\beta_1}\nonumber\\
&\times&
\left(e^{\frac{1}{2\xi}\alpha_1}
\sxi{\alpha_2-\beta_1+\pi i}
-
e^{\frac{1}{2\xi}\alpha_2}
\sxi{\beta_1-\alpha_1+\pi i}
\right)\nonumber\\
&-&\overline{R}(\beta_1-\beta_2)_{1,-1}^{0,0}
\sxi{\alpha_2-\beta_1+\pi i}
\sxi{-\alpha_1+\beta_2+\pi i}
e^{\frac{1}{2\xi}(\alpha_1+\alpha_2
-\beta_1-\beta_2)}\nonumber\\
&-&
\overline{R}(\beta_1-\beta_2)_{1,-1}^{-1,1}
i\sxi{\alpha_1-\beta_1+\pi i}\sxi{
\alpha_2-\beta_1+\pi i}
e^{-\frac{1}{2\xi}\beta_2}
\nonumber\\
&\times&\left.
\left(
e^{\frac{1}{2\xi}\alpha_1}\sxi{\alpha_2-\beta_2+\pi i}
-e^{\frac{1}{2\xi}\alpha_2}
\sxi{\beta_2-\alpha_1+\pi i}
\right)
\right\}.
\end{eqnarray}
Because of antisymmetric relation
$h_\lambda(\alpha)=-h_\lambda(-\alpha)$,
we get the equation (\ref{G(-1,1)})
from the following relation of 
trigonometric function.
\begin{eqnarray}
{\rm Int}(\alpha_1 \alpha_2|\beta_1 \beta_2)_{-1,1}
-{\rm Int}(\alpha_2 \alpha_1|\beta_1 \beta_2)_
{-1,1}=0.\label{te1}
\end{eqnarray}
Other cases are shown as the same manner as the above
case.
We compare integrands directly.

~\\
${\bf \bullet}~N\geq 2$, General Case\\
Next we prove general case.
There exists
the following decomposition
of the integral representation
$G_{2N}
(\beta_1 \cdots \beta_{2N})
_{j_1 \cdots j_{2N}}$.
\begin{eqnarray}
&&\overline{G}_{2N}
(\beta_1,\cdots,
j_s,j_{s+1},\cdots,\beta_{2N})_{j_1,
\cdots,j_s,j_{s+1},\cdots,j_{2N}}\\
&=& \cdots \int_{C}
\frac{d\alpha_1'}{2\pi i}
\cdots
\int_{C}
\frac{d\alpha_{j_s+j_{s+1}+2}'}{2\pi i}
\cdots
J(\alpha_1' \cdots 
\alpha_{j_s+j_{s+1}+2}'|\beta_s,\beta_{s+1})_{j_s,j_{s+1}}
\nonumber\\
&\times&{\rm Sym}(
\cdots, \alpha_1' \cdots 
\alpha_{j_s+j_{s+1}+2}',\cdots|
\cdots, \beta_s,\beta_{s+1},\cdots).\nonumber
\end{eqnarray}
Here $J(\alpha_1' \cdots 
\alpha_{j_s+j_{s+1}+2}'
|\beta_s,\beta_{s+1})_{j_s,j_{s+1}}
$ is the integrand of
$\overline{G}_2(\beta_s,\beta_{s+1})_{j_s,j_{s+1}}$,
\begin{eqnarray}
\overline{G}_2(\beta_s,\beta_{s+1})
_{j_{s},j_{s+1}}
=\int_{C} 
\frac{d\alpha_1'}{2\pi i} \cdots
\int_{C}
\frac{d\alpha_{j_s+j_{s+1}+2}'}{2\pi i}
J(\alpha_1' \cdots  
\alpha_{j_s+j_{s+1}+2}'
|\beta_s,\beta_{s+1})_{j_s,j_{s+1}}.
\end{eqnarray}
Here $\overline{G}_2(\beta_1,\beta_2)
_{j_s,j_{s+1}}$
is formally introduced by using
the equation (\ref{Main})
for not only
$j_s+j_{s+1}=0$
but also $
j_s+j_{s+1}=
\pm1, \pm2$.
The function
$
{\rm Sym}(
\cdots, \alpha_1' \cdots \alpha_{j_s+j_{s+1}+2}',\cdots|
\cdots, \beta_s,\beta_{s+1},\cdots)$
is symmetric with respect to
$\alpha_s' \leftrightarrow
\alpha_t'$ and $
\beta_s
\leftrightarrow \beta_{s+1}$.
Therefore $R$-matrix symmetry
(\ref{Dif1'})
is reduced to the 
following relations of trigonometric functions.
\begin{eqnarray}
&&\sum_{\sigma \in S_{(j_s+j_{s+1}+2)}}
J(\alpha_{\sigma(1)}',
\cdots, \alpha_{\sigma(j_s+j_{s+1}+2)}'
|\beta_{s+1},\beta_s)_{j_{s+1},j_s}\\
&=&
\sum_{\sigma \in S_{(j_s+j_{s+1}+2)}}
\sum_{j_s',j_{s+1}'=\pm1,0}
\overline{R}(\beta_s-\beta_{s+1})
_{j_s,j_{s+1}}
^{j_s',j_{s+1}'
}
J(\alpha_{\sigma(1)}',
\cdots, \alpha_{\sigma(j_s+j_{s+1}+2)}'
|\beta_{s},\beta_{s+1})_{j_s',j_{s+1}'}.
\nonumber
\end{eqnarray}
The above equations 
for $j_s+j_{s+1}=0$,
have appeared
in $N=1$ case.
For $j_s+j_{s+1}=\pm1,\pm2$,
we have to do tedious checking.

\subsection{Cyclicity Condition}
Because of the relation $\rho(\beta)=
\rho(i\lambda-\beta)$,
cyclicity condition (\ref{Dif2}) is 
equivalent to
\begin{eqnarray}
\overline{G}_{2N}(\beta_1,\cdots,
\beta_{2N-1},\beta_{2N}-i\lambda)
_{j_1,\cdots,j_{2N}}=
\overline{G}_{2N}(\beta_{2N},\beta_1,\cdots,
\beta_{2N-1})
_{j_{2N},j_1,\cdots,j_{2N-1}}.
\label{Dif2'}
\end{eqnarray}
${\bullet}~N=1$ Case\\
First we demonstarate
how to prove (\ref{Dif2})
in simple case, $N=1$.
Let us consider the case,
\begin{eqnarray}
\overline{G}_2(\beta_1,\beta_2-i\lambda)_{-1,1}
=
\overline{G}_2(\beta_2,\beta_1)_{1,-1}.
\end{eqnarray}
In the RHS we change the variable
$\alpha_1 \to \alpha_1-i\lambda,
\alpha_2 \to \alpha_2-i\lambda$.
The integrands of the LHS and the
RHS coincides because of the relation,
\begin{eqnarray}
\frac{\varphi(\alpha-i\lambda)}
{\varphi(\alpha)}=
-\frac{\sxi{\alpha-\pi i}}{
\sxi{\alpha+\pi i-\lambda i}}.
\end{eqnarray}
We have to check that the contours for the
LHS and the RHS are the same.
Consider the LHS.
The points
 $\alpha_j=\beta_1+\pi i$
which appear in
poles of the kernel
$\varphi(\alpha_j-\beta_1)$
are actualy not poles,
because there exist zeros from the factor
$
\sxi{-\alpha_1+\beta_1+\pi i}
\sxi{-\alpha_2+\beta_1+\pi i
}$.
Therefore the integral $\int
_{C+i\lambda}
d\alpha_1
\int
_{C+i\lambda}
d\alpha_2
$
can be deformed to
$
\int_{C}d\alpha_1
\int_{C}d\alpha_2
$.
The parameter condition,
\begin{eqnarray}
\lambda < 2\pi(\xi+1),
\end{eqnarray}
ensures that there is no poles in deforming
strips of the integral.
Therefore there exists no difference in integral
contours 
between the LHS and the RHS.
Other cases are proved as the same manner as
this caes.

~\\
${\bullet}$ General $N\geq2$ Case\\
Let us consider general case
(\ref{Dif2'}).
For simplicity
we consider the case $j_{2N}=+1$.
We make the following change of
integration variables :
\begin{eqnarray}
\alpha_{0,2N} \to \alpha_{0,2N}-i\lambda,~~
\alpha_{1,2N} \to \alpha_{1,2N}-i\lambda,~~
{\rm in~the~LHS},
\end{eqnarray}
and
\begin{eqnarray}
\{\alpha_{j,1}\}_{j=0}^{j_{2N}}
\to
\{\alpha_{j,2N}\}_{j=0}^{j_{2N}},
\{\alpha_{j,2}\}_{j=0}^{j_{1}}
\to
\{\alpha_{j,1}\}_{j=0}^{j_{1}},
\cdots,
\{\alpha_{j,2N}\}_{j=0}^{j_{2N-1}}
\to
\{\alpha_{j,2N-1}\}_{j=0}^{j_{2N-1}},
\\
{\rm in~the~RHS}.\nonumber
\end{eqnarray}
After changing variables,
the difference 
between the LHS and the RHS
appears only in the factors
involving variables
$\beta_{2N}, \alpha_{0,2N},
\alpha_{1,2N}$.
The factor of LHS, which involves
the variables
$\beta_{2N}, \alpha_{0,2N},
\alpha_{1,2N}$ is
\begin{eqnarray}
&&g_{j_{2N}}(\{\alpha_{j,2N}\}_{j=\pm1,0})
\prod_{s=1}^{2N-1}
\prod_{j=0}^{j_s}
J_\lambda(\alpha_{j,s}-\beta_{2N}+i\lambda)
\prod_{j=0,+1}
J_\lambda(\beta_{2N}-\alpha_{j,2N})
~
I_\lambda(\alpha_{0,2N}-\alpha_{1,2N})
\nonumber\\
&\times&
\prod_{s=1}^{2N-1}
\prod_{j=0,+1}J_\lambda(\beta_s-\alpha_{j,2N}
+i\lambda)
\prod_{s=1}^{2N-1}
\prod_{j=0}^{j_s}
\prod_{j'=0,+1}I_\lambda(\alpha_{j,s}-
\alpha_{j',2N}+i\lambda),\label{cf1}
\end{eqnarray}
Those of RHS is
\begin{eqnarray}
&&g_{j_{2N}}(\{\alpha_{j,2N}\}_{j=\pm1,0})
\prod_{s=1}^{2N-1}
\prod_{j=0}^{j_s}J_\lambda
(\beta_{2N}-\alpha_{j,s})
\prod_{j=0,+1}J_\lambda
(\beta_{2N}-\alpha_{j,2N})~
I_\lambda(\alpha_{0,2N}-\alpha_{1,2N})
\nonumber\\
&\times&
\prod_{s=1}^{2N-1}\prod_{j=0,+1}
J_\lambda(\alpha_{j,2N}-\beta_s)
\prod_{s=1}^{2N-1}
\prod_{j=0}^{j_s}
\prod_{j'=0,+1}
I_\lambda(\alpha_{j',2N}-\alpha_{j,s}).
\label{cf2}
\end{eqnarray}
Here we have used $I_\lambda(\alpha),
J_\lambda(\alpha)$ in (\ref{f1}), (\ref{f2}).
Using the following property,
\begin{eqnarray}
I_\lambda(\alpha)=I_\lambda(i\lambda-\alpha),
~~J_\lambda(\alpha)=J_\lambda(i\lambda-\alpha),
\end{eqnarray}
we know
that (\ref{cf1}) and (\ref{cf2})
are the same.
We have shown that the integrands of
the LHS and the RHS are the same.
We also have to check 
that the contours for the LHS
and the RHS are the same.
Consider the LHS.
The points
$\alpha_{j,2N}=\beta_{s}+\pi i,
(j=0,+1; s=1,\cdots, 2N-1)$
which appear in poles of the kernel
$\varphi(\alpha_{j,2N}-\beta_{s})$
are actualy not poles,
because there exist zeros from the
factor $
\prod_{s=1}^{2N-1}
\prod_{j=0,+1}
\sxi{-\alpha_{j,2N}+\beta_s+\pi i}
$.
Therefore the integral
$\int_{C+i\lambda}
d\alpha_{0,2N}
\int_{C+i\lambda}
d\alpha_{1,2N}
$ can be deformed to
$\int_{C}
d\alpha_{0,2N}
\int_{C}
d\alpha_{1,2N}$.
The parameter condition
$\lambda<
2\pi(\xi+1)$
ensures that there is no poles in
deforming strips of the integral.
We have proved the equation
(\ref{Dif2'}) for $j_{2N}=+1$.
For $j_{2N}=0,-1$ case
we can show the relation
(\ref{Dif2'})
as the same manner as the case $j_{2N}=+1$.

\subsection{Recursion Relation}
It can be shown that
Recursion relation
(\ref{Dif3}) are consequence of
$R$-matrix symmetry (\ref{Dif1}),
unitarity (\ref{unitary}),
crossing symmetry (\ref{crossing}),
and
\begin{eqnarray}
G_{2N}(\beta_1,\cdots,\beta_{2N-2},
\beta,\beta+3\pi i)
_{j_1,\cdots,j_{2N-2},j,-j}
=C_j
G_{2N-2}(\beta_1,\cdots,\beta_{2N-2})
_{j_1,\cdots,j_{2N-2}}.\label{Dif3'}
\end{eqnarray}
We shall prove (\ref{Dif3'}).
Consider the integral formula (\ref{Main}).
The factor $\rho(\beta_{2N-1}-\beta_{2N})$
has zero at $\beta_{2N}=\beta_{2N-1}+3\pi i$.
We see that
\begin{eqnarray}
\rho(\beta_{2N-1}-\beta_{2N})=
\rho(3\pi i)
\frac{i}{4\xi}\left(
{\rm cos}\left(\frac{\pi}{2\xi}\right)
{\rm sin}\left(\frac{3\pi}{2\xi}\right)
\right)^{-1}
(\beta_{2N}-\beta_{2N-1}-3\pi i)+\cdots.
\end{eqnarray}
The integral may have a pole 
at $\beta_{2N}=\beta_{2N-1}+3\pi i$
because the integral contour
$(-\infty, \infty)$ is pinched by the pole
of the kernel function $\Psi(\{\alpha\}|
\{\beta\})$.
We explain this procedure using
simple example.
For regular function $f(\alpha_1,\alpha_2)$,
the following estimate holds.
\begin{eqnarray}
\int_{C_1}
\frac{d\alpha_1}{2\pi i}
\int_{C_2}
\frac{d\alpha_2}{2\pi i}
f(\alpha_1,\alpha_2)
\frac{1}{(\beta_1-\alpha_1+\pi i)
(\alpha_1-\alpha_2+\pi i)
(\alpha_2-\beta_2+\pi i)}
\nonumber\\
=
\frac{f(\beta_1+\pi i,
\beta_1+2\pi i)}{(\beta_1-\beta_2+3\pi i)}
+{\rm Regular},~~~(\beta_2 \to \beta_1+3\pi i).
\end{eqnarray}
Here the contour $C_1$ is taken along a
path from $-\infty$ to $\infty$
in such a way that
$-\pi<{\rm Im}(\alpha_2-\beta_2)$,
${\rm Im}(\alpha_1-\alpha_2)<\pi$, and
the contour $C_2$ is taken along a
path from $-\infty$ to $\infty$
in such a way that
$-\pi<{\rm Im}(\alpha_1-\alpha_2)$,
${\rm Im}(\beta_1-\alpha_1)<\pi$.
We will check this procedure
for our considering case,
and calculate the residue of $\overline{G}
_{2N}$.
\\
${\bullet}~N=1$ Case\\
First we demonstrate
simple case, $N=1$.
\begin{eqnarray}
{G}_2(\beta_1,\beta_2)_{j,-j},~~
~\beta_2 \to \beta_1+3\pi i.
\end{eqnarray}
The integral $\int_C d\alpha_1$ 
is pinched by the pole of
$\varphi(\alpha_1-\beta_1)$,
and the
integral $\int_C d\alpha_2$ 
is pinched by the pole of
$\varphi(\alpha_2-\alpha_1)$.
Because
$S_2(\omega_1|\omega_1\omega_2)=
\sqrt{}$
The kernel function
$
\varphi(\alpha_1-\alpha_2)
\psi(\alpha_1-\alpha_2)
\prod_{s,t=1,2}
\varphi(\alpha_s-\beta_t)
$ behaves
for $\beta_1 \to \alpha_1+\pi i$,
$\alpha_1 \to \alpha_2+2\pi i$,
$\alpha_2 \to \beta_2+\pi i$.
\begin{eqnarray}
\varphi(2\pi i)^2\psi(\pi i)
\left(
\frac{\sqrt{\frac{\lambda \xi}{2\pi}}}
{iS_2(2\pi|\lambda,2\pi\xi)}\right)^3
\frac{-1}{(\beta_1-\alpha_1+\pi i)
(\alpha_1-\alpha_2+\pi i)
(\alpha_2-\beta_2+\pi i)}+\cdots.
\end{eqnarray}
Note that
\begin{eqnarray}
\varphi(2\pi i)\psi(\pi i)h_\lambda(\pi i)
=\left(2^6 \varphi(0)
{\rm sh}^2\left(\frac{\pi i}{2\xi}\right)
{\rm sh}\left(\frac{3 \pi i}{2\xi}\right)
\right)^{-1}.
\end{eqnarray}
We know
the following is independent of 
the spectral parameter $\beta$.
\begin{eqnarray}
{G}_2
(\beta,\beta+3\pi i)_{j,-j}=C_j,~~(j=\pm1,0),
\end{eqnarray}
where
\begin{eqnarray}
C_1&=&
\frac{
\sqrt{2}
e^{-\frac{\pi i}{2\xi}}}{\xi}
~\sqrt{\frac{\lambda \xi}{\pi}}^3
{\rm sh}^2\left(\frac{\pi i}{\xi}\right)
~
\frac{\rho(3\pi i)}{S_2(2\pi|\lambda,2\pi\xi)^3}
\frac{\varphi(2\pi i)}{\varphi(0)},
\label{C1}\\
C_0&=&
\frac{
\sqrt{2}e^{-\frac{\pi i}{\xi}}}{i \xi}
~\sqrt{\frac{\lambda \xi}{\pi}}^3
{\rm sh}\left(\frac{\pi i}{2\xi}\right)~
\frac{\rho(3\pi i)}{S_2(2\pi|\lambda,2\pi\xi)^3}
\frac{\varphi(2\pi i)}{\varphi(0)},
\label{C0}\\
C_{-1}&=&
\frac{\sqrt{2}
e^{-\frac{3\pi i}{2\xi}}}
{\xi}
~\sqrt{\frac{\lambda \xi}{\pi}}^3
{\rm sh}^2
\left(\frac{\pi i}{\xi}\right)~
\frac{\rho(3\pi i)}{S_2(2\pi|\lambda,2\pi\xi)^3}
\frac{\varphi(2\pi i)}{\varphi(0)}.\label{C-1}
\end{eqnarray}

~\\
${\bullet}$ General $N\geq2$ Case\\
We show that general $N\geq 2$ case
is reduced to $N=1$ case.
Consider the integrand of the 
following integral
formula.
\begin{eqnarray}
G_{2N}(\beta_1,
\cdots,\beta_{2N-2},
\beta_1',\beta_2')_{j_1\cdots j_{2N-2},j,-j}.
\end{eqnarray}
Let us set integral variables,
$$\{\alpha_1',\alpha_2'\}
=\{\{\alpha_{k,2N-1}\}_{k=0}^j,
\{\alpha_{k,2N}\}_{k=0}^{-j}
\}.$$
The factor interacting both 
$\alpha_{j,s}, (s=1,\cdots,2N-2;j=\pm1,0)$
and
$\beta_1',\beta_2',\alpha_1',\alpha_2'$
is given by
\begin{eqnarray}
&&\prod_{s=1}^{2N-2}
\rho(\beta_s-\beta_1')
\rho(\beta_s-\beta_2')
J_\lambda(\beta_s-\alpha_1')
J_\lambda(\beta_s-\alpha_2')\nonumber
\\&\times&
\prod_{s=1}^{2N-2}
\prod_{j=0}^{j_s}
J_\lambda(\alpha_{j,s}-\beta_1')
J_\lambda(\alpha_{j,s}-\beta_2')
I_\lambda(\alpha_{j,s}-\alpha_1')
I_\lambda(\alpha_{j,s}-\alpha_2').
\label{separate}
\end{eqnarray}
The above factor (\ref{separate})
simplifies for
$\beta_2'\to \alpha_2'+\pi i,
\alpha_2'\to \alpha_1'+\pi i,
\alpha_1' \to \beta_1'+\pi i$,
\begin{eqnarray}
(\ref{separate}) \to 2^{-28(N-1)},
\end{eqnarray}
because of the following relations.
\begin{eqnarray}
\rho(\beta)\rho(\beta-3\pi i)
J_\lambda(\beta-\pi i)
J_\lambda(\beta-2\pi i)=2^{-2},
\\
J_\lambda(\beta)J_\lambda(\beta-3\pi i)
I_\lambda(\beta-\pi i)
I_\lambda(\beta-2\pi i)=2^{-12}.
\end{eqnarray}
This simplification teaches us
that general $N\geq 2$ case
is reduced to
the simplest $N=1$ case.
We have proved recursion relation 
(\ref{Dif3}).

\section{Supporting Argument}
The purpose of this section
is to give a supporting argument
on physical meaning of our integral representation.
Let us consider the integral formula
for $\lambda=3\pi$.
In this case following simplifications occure.
\begin{eqnarray}
\rho(\beta)=
\frac{S_2(-i\beta+2\pi|6\pi,2\pi\xi)
S_2(-i\beta+3\pi|6\pi,2\pi\xi)
}{
S_2(-i\beta+2\pi\xi|6\pi,2\pi\xi)
S_2(-i\beta+2\pi\xi+\pi|6\pi,2\pi\xi)
},
\end{eqnarray}
and
\begin{eqnarray}
\psi(\alpha)=\varphi(\alpha)\times
\frac{1}{4 \sxi{-\alpha-2\pi i} 
\sxi{\alpha-2\pi i}}.
\end{eqnarray}
For the case $\lambda=3\pi$,
our integral formula 
(\ref{Main}) gives
a conjectural formula
for the Izergin-Korepin model
\cite{IK}.
Hou et al.\cite{HYZ, M} considered 
the correlation functions
for the Izergin-Korepin model
for massive regime $-1<{q}<0$,
within the framework
of representation theory
of $U_q(A_2^{(2)})$.
They
gave free field realizations of
the vertex operators,
and realized integral representation
of the correlation functions,
as the trace of the vertex operators.
In the limiting case,
our integral representation
reproduce
the correlation function
for the Izergin-Korepin model
at critical point $q=-1$,
which was derived by Hou et al.\cite{HYZ, M}.
This gives a supporting argument
that our integral representation (\ref{Main})
gives a conjectural formula of the correlation
function of the Izergin-Korepin model
at massless regime $|q|=1$.

In the special case
where
$\lambda=3\pi$ and $\xi \to \infty$,
the auxiliary functions
$g_{k}(\{\alpha_j\}_{j=-1}^k), (k=\pm1,0)$ tend to
$\widetilde{g}_{k}(\{\alpha_j\}_{j=-1}^k), (k=\pm1,0)
$ given by
\begin{eqnarray}
&&\widetilde{g}_{-1}(\{\alpha_{-1}\})=1,\\
&&\widetilde{g}_{0}(\{\alpha_{-1}, \alpha_0\})=
\frac{2\pi i}{\alpha_0-\alpha_{-1}+\pi i},\\
&&\widetilde{g}_{1}(\{\alpha_{-1}, \alpha_0,
\alpha_1\})=
\frac{2\pi(2\alpha_{-1}-\alpha_0-\alpha_1)}
{(\alpha_0-\alpha_{-1}+\pi i)
(\alpha_1-\alpha_{-1}+\pi i)}.
\end{eqnarray}
The intergal representation 
$G_{2N}$,
(\ref{Main}) becomes as follows.
\begin{eqnarray}
&&\prod_{1\leq s <t \leq 2N}
\frac{1}{\Gamma\left(
\frac{i(\beta_s-\beta_t)}{6\pi}+1
\right)
\Gamma\left(\frac{i(\beta_s-\beta_t)}{6\pi}+\frac{5}{6}
\right)
\Gamma\left(
\frac{i(-\beta_s+\beta_t)}{6\pi}+\frac{1}{2}
\right)
\Gamma\left(
\frac{i(-\beta_s+\beta_t)}{6\pi}+\frac{1}{3}
\right)}\nonumber\\
&\times&
\prod_{s=1}^{2N}\prod_{j=0}^{j_r}
\int_{-\infty}^\infty
\frac{d\alpha_{j,s}}{2\pi i}~
\widetilde{\Phi}(\{\alpha\}|\{\beta\})~
\prod_{s=1}^{2N}\widetilde{g}_{j_s}
(\{\alpha_{j,s}\}_{j=-1}^{j_s})
\nonumber\\
&\times&
\prod_{s=1}^{2N}
\prod_{j=0,1}
\prod_{t \in {\cal A}_j
\atop{s\leq t}}
(-\beta_s+\alpha_{j,t}+\pi i)
\prod_{t \in {\cal A}_j
\atop{t<s}}
(-\alpha_{j,t}+\beta_s+\pi i)
\nonumber\\
&\times&
\prod_{j=0,1}
\prod_{s,t \in {\cal A}_j
\atop{s<t}}
\frac{\alpha_{j,s}-\alpha_{j,t}-\pi i}{
\alpha_{j,s}-\alpha_{j,t}-2\pi i 
}
\prod_{s \in {\cal A}_0,t \in {\cal A}_1
\atop{s\leq t}}
\frac{\alpha_{0,s}-\alpha_{1,t}-\pi i}
{\alpha_{0,s}-\alpha_{1,t}-2\pi i
}
\prod_{s \in {\cal A}_0,t \in {\cal A}_1
\atop{s>t}}
\frac{
\alpha_{1,t}-\alpha_{0,s}-\pi i
}{
\alpha_{1,t}-\alpha_{0,s}-2\pi i 
}
\nonumber\\
&\times&
\prod_{j=0,1}
\prod_{s,t \in {\cal A}_j
\atop{s<t}}
{\rm sh}\left(\frac{1}{3}(
\alpha_{j,s}-\alpha_{j,t})\right)
\prod_{s \in {\cal A}_0,t \in {\cal A}_1
\atop{s\leq t}}
{\rm sh}\left(
\frac{1}{3}(\alpha_{0,s}-\alpha_{1,t})\right)
\prod_{s \in {\cal A}_0, t \in {\cal A}_1
\atop{s>t}}
{\rm sh}\left(\frac{1}{3}(
\alpha_{1,t}-\alpha_{0,s})\right).\nonumber\\
\end{eqnarray}
Here we have set the kernel function
$\widetilde{\Phi}(\{\alpha\}|\{\beta\})$ by
\begin{eqnarray}
\widetilde{\Phi}(\{\alpha\}|\{\beta\})&=&
\prod_{s=1}^{2N}\prod_{j=0,1}
\prod_{t \in {\cal A}_j}
\widetilde{\varphi}(\beta_s-\alpha_{j,t})
\prod_{s \in {\cal A}_0, t \in {\cal A}_1}
\widetilde{\varphi}(\alpha_{0,s}-\alpha_{1,t})
\widetilde{\phi}(\alpha_{0,s}-\alpha_{1,t})\nonumber\\
&\times&
\prod_{j=0,1}
\prod_{s,t \in {\cal A}_j
\atop{s<t}}
\widetilde{\varphi}(\alpha_{j,s}-\alpha_{j,t})
\widetilde{\phi}(\alpha_{j,s}-\alpha_{j,t}),
\end{eqnarray}
where
\begin{eqnarray}
\widetilde{\varphi}(\alpha)=
\Gamma\left(\frac{i\alpha}{3\pi}+\frac{1}{3}\right)
\Gamma\left(
-\frac{i\alpha}{3\pi}+\frac{1}{3}
\right),~
\widetilde{\phi}(\alpha)=
\frac{1}{\displaystyle
\Gamma\left(\frac{i\alpha}{3\pi}+\frac{2}{3}\right)
\Gamma\left(
-\frac{i\alpha}{3\pi}+\frac{2}{3}
\right)
}.
\end{eqnarray}
In the limiting case, our integral representation 
reproduce those of the correlation function
at critical point $q=-1$ \cite{HYZ}.

Recently Kitanine, Maillet, Terras
\cite{KMT} derived integral representation
of the correlation function for the six-vertex model
($A_1^{(1)}$ symmetry) 
by means of the quantum inverse scattering method.
Their method is avairable for both massive and massless
regime.
To construct $A_2^{(2)}$
analogue of the paper \cite{KMT}
and reproduce the integral representation
(\ref{Main})
is interesting problem.


\begin{thebibliography}{99}
\bibitem{HYZ}B.Hou, W.Yang and Y.Zhang:
The twisted quantum affine algebra $U_q(A_2^{(2)})$
and correlation functions of the Izergin-Korepin model,
{\it Nucl.Phys.}{\bf B556},
485-504, (1999).
\bibitem{M}Y.Matsuno: Thesis (Kyoto University,
superviser : M.Jimbo),
1997.
\bibitem{J}N.Jing :
Twisted vertex representations of quantum affine
algebras,
{\it Invent.Math.}{\bf 102},
663-690, (1990).
\bibitem{D}V.Drinfeld:
New realizations of Yangians and quantum affine algebras,
{\it Sov.Math.Dokl.}
{\bf 36},212-216, (1988).
\bibitem{IK}
A.Izergin and V.Korepin:
The invese scattering method approach to the quantum
Shabat-Mikhailov model,
{\it Commun.Math.Phys.}{\bf 79}, 303-316, (1981).
\bibitem{S}
F.Smirnov:
Form factors in completely integrable
models of quantum field theory,
Advanced Series in Mathematical Physics, {\bf 14},
World Scientific, 1992.
\bibitem{JM1}M.Jimbo and T.Miwa :
Quantum KZ equation with $|q|=1$ and correlation functions
of the XXZ model in the gapless regime,
{\it J.Phys.}{\bf A29}, 2923-2958, (1996).
\bibitem{JKM}M.Jimbo, H.Konno and T.Miwa :
Massless XXZ model and degeneration
of the elliptic algebra ${\cal A}_{q,p}(\widehat{sl_2})$,
{\it Deformation theory and Symplectic Geometry},
(Kluwer), (1997).
\bibitem{K1}T.Kojima : The 19 Vertex Model at
critical regime $|q|=1$,
{\it Int.J.Mod.Phys.}{\bf A16}, 1559-1578, (2001).
\bibitem{KY}T.Kojima and S.Yamasita :
The critical $A_{n-1}^{(1)}$ chain,
{\it J.Phys.}{\bf A34}:Math.Gen.,1181-1201, (2001).
\bibitem{MT}T.Miwa and Y.Takeyama :
The integral formula for the solutions of the quantum
Knizhnik-Zamolodchikov equation associated with
$U_q(\widehat{sl_n})$ for $|q|=1$,
{\it Proc.SIDE III Meeting (Sabaudia)}, (1999).
\bibitem{TV}V.Tarasov and A.Varchenko :
Geometry of $q$-hypergeometric functions
as a bridge between Yangians and quantum affine
algebras,
{\it Invent.Math.}
{\bf 128}, 501-588, (1997).
\bibitem{KMT}
N.Kitanine,J.M.Maillet and V.Terras :
Correlation functions
of the XXZ Heisenberg spin-$\frac{1}{2}$
chain in a magnetic field,
{\it Nucl.Phys.}{\bf B567},no.3, 554-584, (2000).
\end{thebibliography}
\end{document}